\documentstyle[aps,prb,preprint,epsfig]{revtex}
\begin{document}
\draft
\newcommand{\ccc}{{\c C}a{\u g}{\i}n}
\title{Vacancy Formation Energy at High Pressures in Tantalum}
\author{Sonali Mukherjee $^{(1,2)}$, R.~E. Cohen $^{(1,2)}$ and O{\u g}uz G{\"u}lseren $^{(1,3,4)}$}
\address{$^{(1)}$ Geophysical Laboratory and Center for High Pressure Research\\ 
Carnegie Institution of Washington,
5251 Broad Branch Road, N.W. Washington, DC 20015\\}
\address{$^{(2)}$ Seismological Laboratory, California Institute Of Technology, Pasadena, CA 91125} 
\address{$^{(3)}$ NIST Center for Neutron Research,
National Institute of Standards and Technology,
Gaithersburg, MD 20899}
\address{$^{(4)}$Department of Materials Science,
University of Pennsylvania, Philadelphia, PA 19104}
\date{\today}
\maketitle
\begin{abstract}
We have computed the variation of the vacancy formation energy 
with pressure 
for Ta. Total energy calculations were performed for 16 and 54 atom
supercells using a mixed basis pseudopotential
method which uses pseudo-atomic orbitals and low 
energy plane waves as the basis set,
within density functional theory (LDA and GGA). 
The vacancy formation energy is found to 
increase from 2.95 eV at ambient pressures to 6.96 eV at 300 GPa, 
and the vacancy formation volume decreases 
from being 53.4$\%$ of the bulk volume per atom at ambient 
pressure to 19.6$\%$ at 300 GPa, for a 54 atom supercell.
The structural relaxation energy is found to   
increase with pressure from 14$\%$ of the vacancy formation energy 
at ambient pressure to 16$\%$ at 300 GPa.

\end{abstract}
\section{Introdution}
\label{sec:Intro}
Vacancies control a number of mechanical and kinetic processes in metals, such as
dislocation climb, self-diffusion etc. Mechanical properties of metals,
are determined in part by finite temperature vacancy concentrations.
Thus the material specific vacancy formation energy, which determines the 
finite temperature vacancy concentration,
serves as an important parameter in understanding 
the mechanical behavior of the metal.
Due to its high thermal, mechanical and chemical stability,
Ta is an important technological material and also a good candidate for 
an internal pressure standard for high pressure experiments. 
We determine the vacancy formation energy 
and its pressure dependence for bcc Ta. 

Direct experimental 
measurement of the pressure dependence of the vacancy formation energy 
would be extremely difficult, making {\em ab-initio} calculations 
particularly useful.\cite{cynn} 
Moreover, the {\em ab-initio} calculation results
can be used to validate classical potentials which can then be used for large 
scale simulations required for understanding processes 
like dislocation motion, fractures etc, 
which occur at much larger length scales as compared to atomic length scales.  

\section{Methodology}
\label{sec:Method}
\begin{figure}
\centerline{\epsfig{file=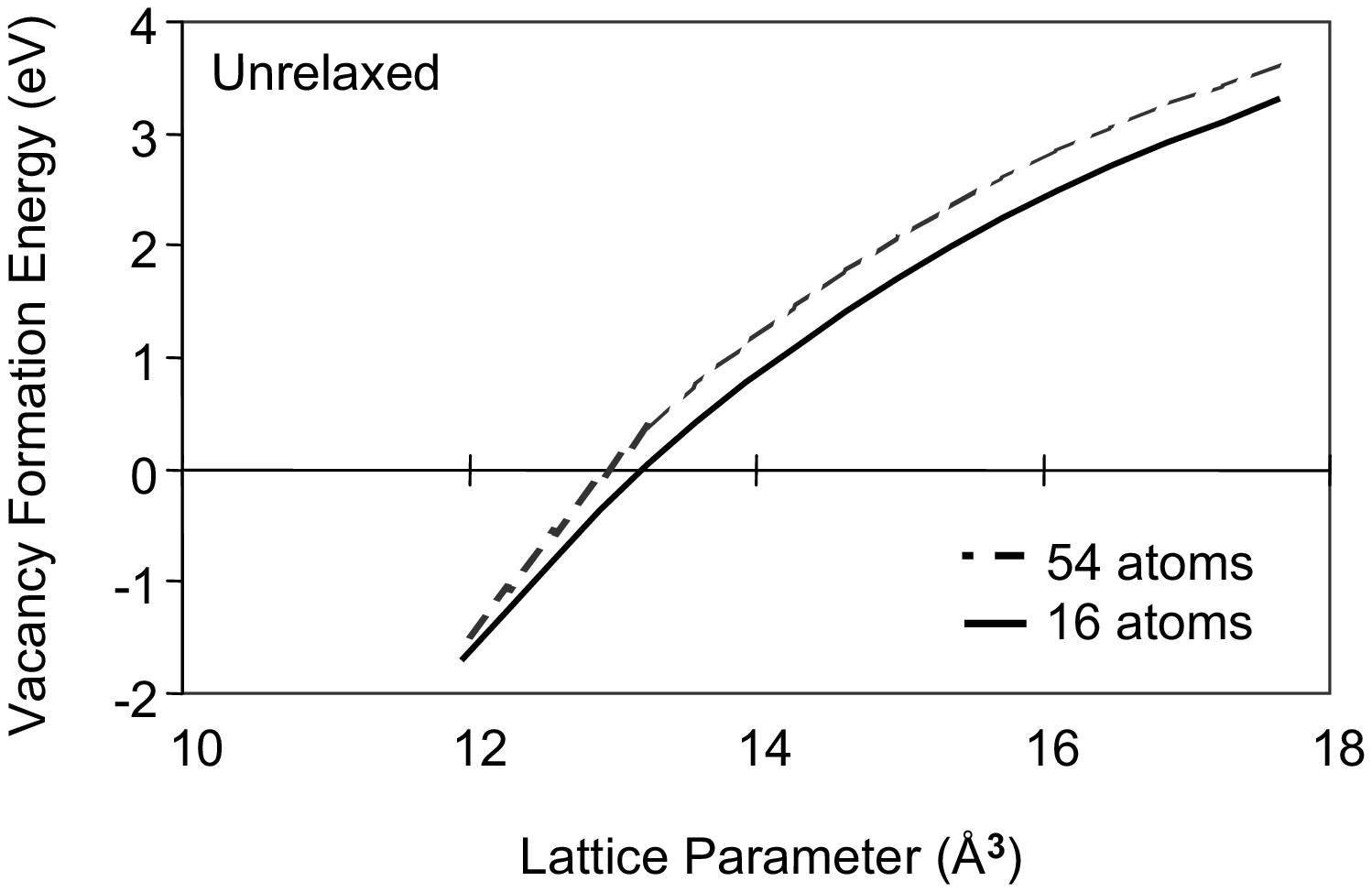,width=3.5in}}
\caption{\label{fig1}
Plot of unrelaxed vacancy formation energy with respect to pressure for
the 16 and 54 atom supercells. As pressure increases
the unrelaxed vacancy formation energy becomes negative indicating the 
importance of relaxation at high pressures.}
\end{figure}

The pseudopotential method within density functional theory (LDA\cite{LDA}
and GGA \cite{GGA})
was used to compute the total energies of the 16 and 54 atom supercells with and 
without a vacancy. 
We generated a 
non-local, norm-conserving  Troullier-Martins \cite{trou} semi-relativistic
pseudopotential. 
We have used a mixed-basis approach which uses pseudo-atomic orbitals 
and a few low-energy plane waves as the basis set,\cite{method1} 
with a cut-off of 60 eV. The cut-off for the plane waves used to 
expand the potential is 550 eV.  
A special k-point mesh\cite{monpack} was used, giving the number of
k-points used in the irreducible wedge of the BZ for the cubic symmetry is 
35(10) for N=16(54), where N is the number of atoms in the supercell.
Convergence of the vacancy formation energy of up to $0.001$ eV 
was achieved with respect to k-point sampling. 
Structural relaxation was performed using analytic forces. 
Relaxation was considered complete when forces on the 
ions were less than $0.001$ (eV/\AA). 
The volume relaxed vacancy formation energy at a given pressure P, is obtained by comparing 
the total energy of a supercell with N atoms $E_{tot}(N,P)$,
with that of a supercell with N-1 atoms and a vacancy 
at the same pressure P, $E_{tot}(N-1,P)$,
\begin{equation}
\label {eq:evac}
E_{vac}(P) = E_{tot}(N-1,P) - (N-1/N) E_{tot}(N,P).
\end{equation}
and similarly for the enthalpy $H = E + P V$. $E$, $P$, and $H$ are obtained 
by fitting the total energy results versus volume to the Vinet equation of 
state.\cite{vinet,coheneta}  
In a bulk system, removal of one atom would not change the bulk pressure. However, for tractable supercell 
sizes 
creation of a 
vacancy in the supercell can change the pressure dramatically, 
especially at high pressure.
Thus relaxation of the system with the vacancy is essential for obtaining sensible 
vacancy formation energies. This is evident from 
Fig.~\ref{fig1} where the unrelaxed vacancy formation energy is plotted 
for the 16 and 54 atom supercells. Notice that at high pressure (smaller lattice constant), 
the vacancy formation energy becomes negative, which is unrealistic.
Notice that the vacancy formation energy becomes
negative at a lower pressure (larger lattice constant) for the 16 atom supercell  
compared to the 54 atom supercell. 
This can be understood as a supercell size effect. In a real crystal
most of the atoms are not displaced from their positions with the 
introduction of a vacancy, 
most atoms are far from the 
vacancy, and the lattice constant of the 
system remains unchanged. As the supercell size decreases, more and more  
atoms undergo displacement because of the vacancy resulting 
in a change of lattice constant and volume relaxation.

\section{Results}
\label{sec:Result}
Our ambient pressure vacancy formation energy  
is in good agreement with other computations (Table I).
Also note that, for the 
54 atom supercell the fully relaxed vacancy formation energies
are within ${3\%}$ of experiment at zero pressure.
The total energy of the
54 atom supercell with and without a vacancy for different system
volumes is tabulated in Table II.

Volume relaxed Ta vacancy formation energies increase 
with pressure for the 16 and 54 atom supercells 
(Fig.~2a). 
The value of the vacancy formation energy is lower for the 
16 atom supercell than the 54 atom
supercell at all pressures. The difference in the vacancy formation energy for the 16 and 54 
atom supercells increases with pressure from 10$\%$ at ambient pressures to 15$\%$ at 300 GPa.

The volume relaxed vacancy formation energy is lower for GGA 
compared to LDA at all pressures 
(Fig.~2b).
The difference between LDA and GGA increases    
slightly with pressure, (from 6$\%$ at ambient pressures  
to 9$\%$ at 300 GPa). We also find that the vacancy formation energy using PBE GGA  
is essentially the same as that obtained by using PW91 GGA.

The fully relaxed (volume and structural relaxation)
and the volume relaxed vacancy formation energies of Ta 
are compared for the 16 and 54 atom supercells (Figs.~2c and 2d). 
For the 54 atom supercell the fully(only volume) relaxed 
vacancy formation energy increases from 2.95(3.26) eV 
at ambient pressure to 6.96(8.16) eV at 300 GPa.
The structural relaxation energy which is the difference between the  
fully relaxed and the volume relaxed vacancy formation energies
is sensitive to the supercell size. It increases with the size of the 
supercell.
At ambient pressure the structural relaxation energy for the 54 atom
supercell is $14\%$ of the vacancy formation energy compared to 
$1\%$ for the 16 atom supercell. 
The high relaxation energy of $14\%$ for the 54 atom supercell at ambient pressures
has been obtained in previous calculations.\cite{will}
The pressure dependence of the relaxation
energy reduces with supercell size. 
The structural relaxation energy
increases from $1\%$ to $10\%$ for the 16 atom supercell compared to
$14\%$ to $16\%$ for the 54 atom supercell with increase in pressure
from ambient to 300 GPa.     
It is important to check the convergence of the vacancy formation energy with respect to the
supercell size. For small supercells introduction of a vacancy in the system
leads to an unreasonably high vacancy concentration which can lead
to drastic change in the elastic properties of the system. 
Moreover, inter-vacancy
interactions in the smaller supercell may alter the vacancy formation energy.
In order to test the convergence with respect to system size,
we compare our results with those of 
molecular dynamics simulations with a periodic cell of 1458 Ta atoms.
\cite{md}
The potential used in the simulations was fitted
to our zero pressure vacancy formation energies in the same sized supercell.
Our fully relaxed vacancy formation energy for the 54 atom supercell
is within $5\%$ of the molecular dynamics results for zero pressure.
The convergence indicates that in the 54 atom supercell
vacancy-vacancy interactions are reduced to
a negligible level, thus allowing us to calculate vacancy
formation energies for an isolated
vacancy. 
The difference between our results and the molecular dynamics 
results increases with pressure to $20\%$ at 300 GPa. 
This difference can be 
attributed to the difference in the  
atomic relaxations 
obtained from the interatomic potential
used in the molecular dynamics simulations 
which are smaller compared to our 
calculations. \cite{md}

The vacancy
formation enthalpy is the important quantity at non-zero pressures. 
The vacancy formation enthalpy increases with pressure similar 
to the vacancy formation energy, but its rate of increase is slower
and it saturates at high pressure (Fig.~\ref{fig3}). This saturation behavior 
is due to decreasing vacancy volume and increasing vacancy formation
energy with increasing pressure. 

Volume relaxation results in displacement of all atoms towards the vacancy with
a consequent reduction in the volume of the supercell. Structural relaxation 
results in oscillatory displacement of atoms around the vacancy with the nearest 
neighbor moving towards the vacancy 
and the successive neighbors moving alternately 
away (negative displacement) and towards (positive displacement), the vacancy. 
The oscillatory atomic displacement around the vacancy due to structural 
relaxation has also been observed   
in Mo at ambient pressures.\cite{xu}   
In the context of a real crystal, the partitioning of the atomic displacement  
due to volume and structural relaxation is rather artificial; probably a 
more relevant quantity
would be the total atomic
displacement due to both volume and structural relaxation. 
The total atomic displacement retains the oscillatory nature in Ta, but all the  
atomic displacements are towards the vacancy (positive) (Fig.~4a).   
The atomic displacement due to volume relaxation enhances the positive
displacement due to structural relaxation and reduces the negative displacement.
Even the furthest neighbor of the vacancy  
has non-zero displacement as  
it includes the effects of both volume and structural relaxations.
With increasing pressure, all the atomic displacements become increasingly
positive, i.e., towards the vacancy. 
The pressure dependence of the displacements for the 
first and second nearest neighbor of the vacancy 
is shown in Fig.~4b. One can see that the 
pressure dependence of the  
first nearest neighbor displacement is negligible 
beyond 100 GPa.
In contrast, the displacement of the second nearest neighbor towards the
vacancy increases
continuously with increasing pressure. 

The vacancy formation volume ($\Omega_{vac}^{f}$), is given by,  
\begin{equation}
\Omega{^{f}}_{vac}(P)=\Omega_{0}(P) - ( \Omega_{N}(P) - \Omega_{N-1}(P)),
\end{equation}
where, $\Omega_{N}=N\Omega_{0}$. $\Omega_{N}$ is the volume of N atom supercell
and $\Omega_{N-1}$ is the volume of the relaxed N-1 atom supercell with a vacancy.
At ambient pressure $\Omega_{vac}^{f}$ is $53.4\%$ of the volume of a single atom in bulk Ta.
Our vacancy formation volume agrees well with other ambient pressure calculations
for Ta.\cite{satta,moriarty} 
Amongst the bcc transition metals Ta has the highest vacancy formation volume 
after W.\cite{moriarty} 
Note that with relaxation $\Omega_{N-1}$ decreases thus reducing the vacancy formation
volume compared to the volume of a single atom $\Omega_{0}$.   
The vacancy formation volume is smaller than   
the volume of a single atom at the same pressure (Fig.~\ref{fig5}). 
As the pressure increases 
the displacement of the atoms towards the vacancy increases, thus 
reducing the vacancy formation volume compared to a single atom at the same pressure.
Fig.~\ref{fig5} shows the decreasing trend of the vacancy formation volume with pressure.
Note that the vacancy formation volume reduction rate appears to approach 
saturation at higher pressures. 
The difference in the vacancy formation volume
due to structural relaxation increases with pressure from being $23.6\%$ at ambient pressure
to $49.5\%$ at 300 GPa for the 54 atom cell.

\section{Discussion}

Transition metals tend to have the highest vacancy formation energies compared
to other elemental metals
because of the presence of strong angular forces
arising from d-state interactions.
Amongst the transition metals, Ta has the second-highest vacancy formation energy
at ambient pressure; W having the highest vacancy formation energy.\cite{moriarty}  
The Ta vacancy formation energy is almost three times
as that of a typical elemental metal like Cu.

Vacancies play an important role in controlling the 
rheology of the material. Plastic deformation 
under an applied stress is due to dislocation motion in 
the system. The strain rate $d\epsilon/dt$ in the quasi-steady-state
regime can be expressed as:
\begin{equation}
d\epsilon/dt = \rho b v,
\end{equation}
where $\rho$ is the density of mobile dislocations, b is the Burgers vector
and v is the average velocity of the dislocations.\cite{poirier}  
A dislocation usually glides until its motion is obstructed by 
impurities, interstitials or is entangled by other dislocations.
After overcoming these obstacles  
the dislocation again glides till it encounters the next obstacle.
So the average dislocation velocity v can be expressed as:
\begin{equation}
v = \delta L/(t_g+t_o),
\end{equation}
where $\delta L$ is the distance traveled by the dislocation,
$t_g$ is the time for which it glides and $t_o$ is the time  
taken to overcome the obstacle. Dislocations can overcome the
obstacle by climbing 
to a plane normal to its glide plane. 
The dislocation climb occurs by self-diffusion of atoms 
with the aid of vacancies.\cite{gir} 
Vacancies also aid 
in the unentanglement of dislocations.  
Thus $t_o$ in the above equation is determined by the vacancy 
concentration. The high vacancy formation energy in Ta will
translate to low probability of dislocation climb and unentanglement
of dislocations. This would result in lower dislocation velocity in Ta
and consequently lower plastic flow and higher yield strength,\cite{weir} 
compared to a typical metal like Cu, especially at high pressure.
Vacancies also cause plastic flow at high temperatures and low stress by
diffusion creep. In diffusion creep in the presence of non-hydrostatic 
stress there arises a stress field in the crystal. 
This in turn results in a flux of vacancies in the crystal causing shear 
deformation.\cite{poirier} The high vacancy formation energy in Ta would reduce the 
probability of diffusion creep even at high temperatures.

In the bcc transition elemental metals, at ambient pressure,
there is a strong
correspondence between
the vacancy formation energy $E_v$, and the melting temperature $T_m$,
(see Fig.~\ref{fig6}). The straight line shows that for all these elements the 
dimensionless quantity, $E_v/k_BT_m$, is similar and is given by 
the slope of the straight line;  
$k_B$ being the Boltzmann constant.
The number of vacancies (n), in a crystal with N atoms,
at a finite temperature (T),   
is determined by the vacancy formation energy ($E_v$);
\begin{equation}
n/N=e^{(-E_v/k_BT)}.
\end{equation}
Thus, similar values of $e^{-E_v/k_BT_m}$ 
indicate that all these metals have 
similar vacancy concentration close to their melting temperatures or 
that the vacancy formation energy and the melting temperature scale 
similarly with the atomic interaction strength.
This indicates that vacancies might play an important role in the
melting of the elemental metals.
Large amplitude atomic vibration and thermal creation of vacancies
have been known to destabilize the crystal and induce melting.\cite{mol,reilly} 
The vacancy concentration for these metals as estimated from the
slope of the straight line is $6.21$x$10^{-4}$.
The close packed elemental metals also exhibit similar correspondence between 
their vacancy formation energies and their melting temperatures.\cite{kittel}

To summarize, our computations show that the mixed basis pseudopotential method
has the potential of giving correct vacancy formation
energies.    
We find that the volume relaxation energies become increasingly important 
with pressure but their importance reduces with increasing supercell
size. On the other hand structural relaxation energies 
increase with the supercell size. 
We find that the difference between LDA and GGA results do not change appreciably
with pressure. The vacancy formation energies were found
not to depend on the kind of GGA exchange correlation used, PBE or PW91. 
The dependence on pressure of the vacancy formation enthalpy 
was found to be different  
compared to the vacancy formation energy. After an initial
increase the enthalpy
saturates with pressure whereas the vacancy formation energy 
increases with pressure without saturation.

\acknowledgments{
This work was supported by DOE ASCI/ASAP subcontract B341492 to
Caltech DOE W-7405-ENG-48. Computations were performed on the Cray SV1
at the Geophysical Laboratory, supported by NSF grant EAR-9975753
and the W.\ M.\ Keck Foundation.  We thank T. \ccc, W.A. Goddard, III, H. Fu 
and A. Strachan, for helpful discussions.}

\begin{figure}
\centerline{\epsfig{file=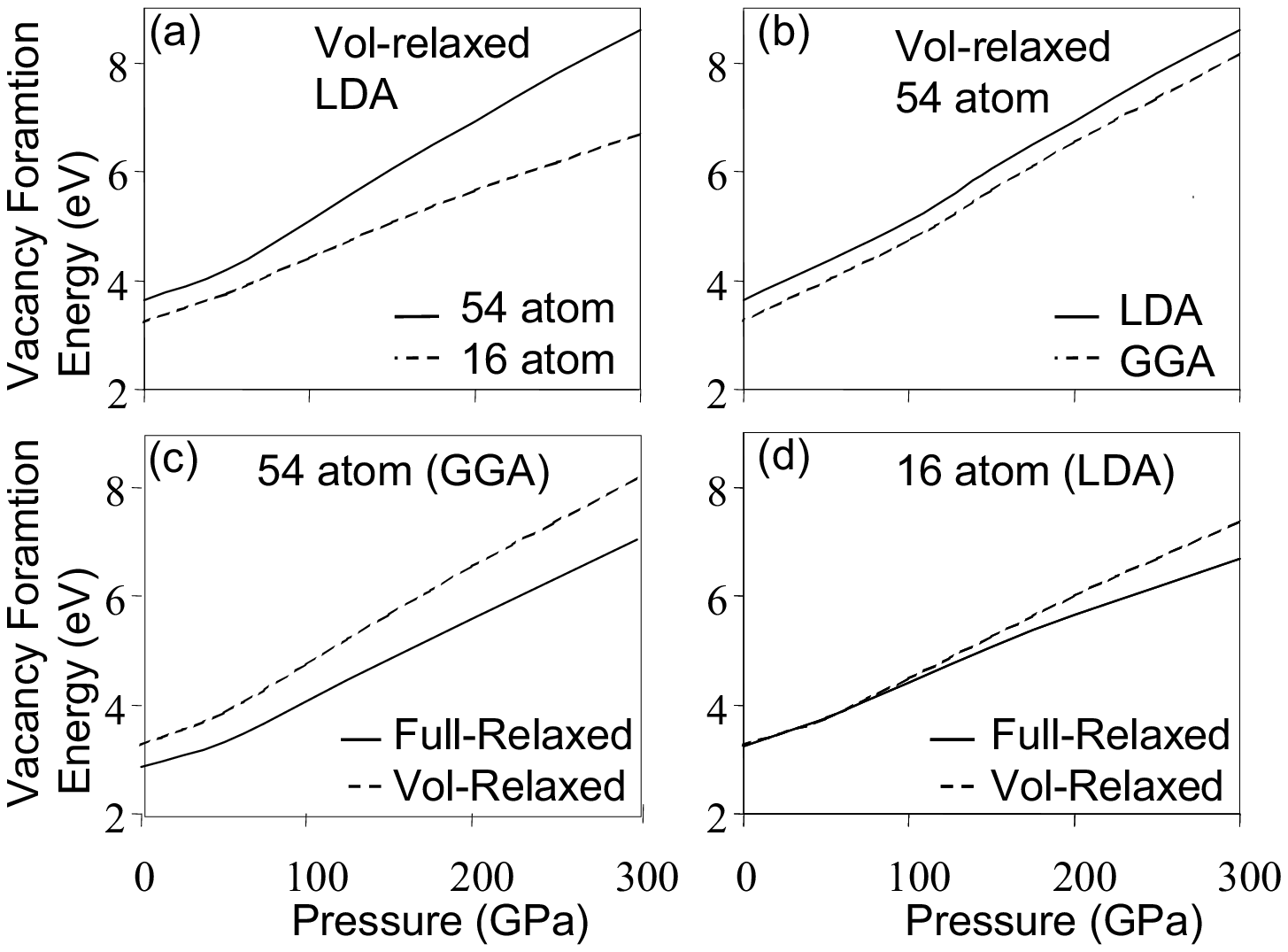,width=4.5in}}
\caption{\label{fig2}
Dependence of the vacancy formation energy $E_{vac}$ on supercell size, 
exchange correlation
and structural relaxations. 
(a)Plot of the volume relaxed vacancy formation energies with respect to pressure for
the 16 and 54 atom supercells. As pressure increases 
the difference in $E_{vac}$ for the two supercell sizes increases.  
(b)Plot of volume relaxed vacancy formation energies with respect to pressure
for the 54 atom supercell for LDA and GGA.
(c)Comparison of volume relaxed and fully relaxed vacancy formation energies
for the 54 atom supercell. (d) Comparison of volume relaxed and fully relaxed vacancy formation energy
for the 16 atom supercell.}
\end{figure}

\begin{figure}
\centerline{\epsfig{file=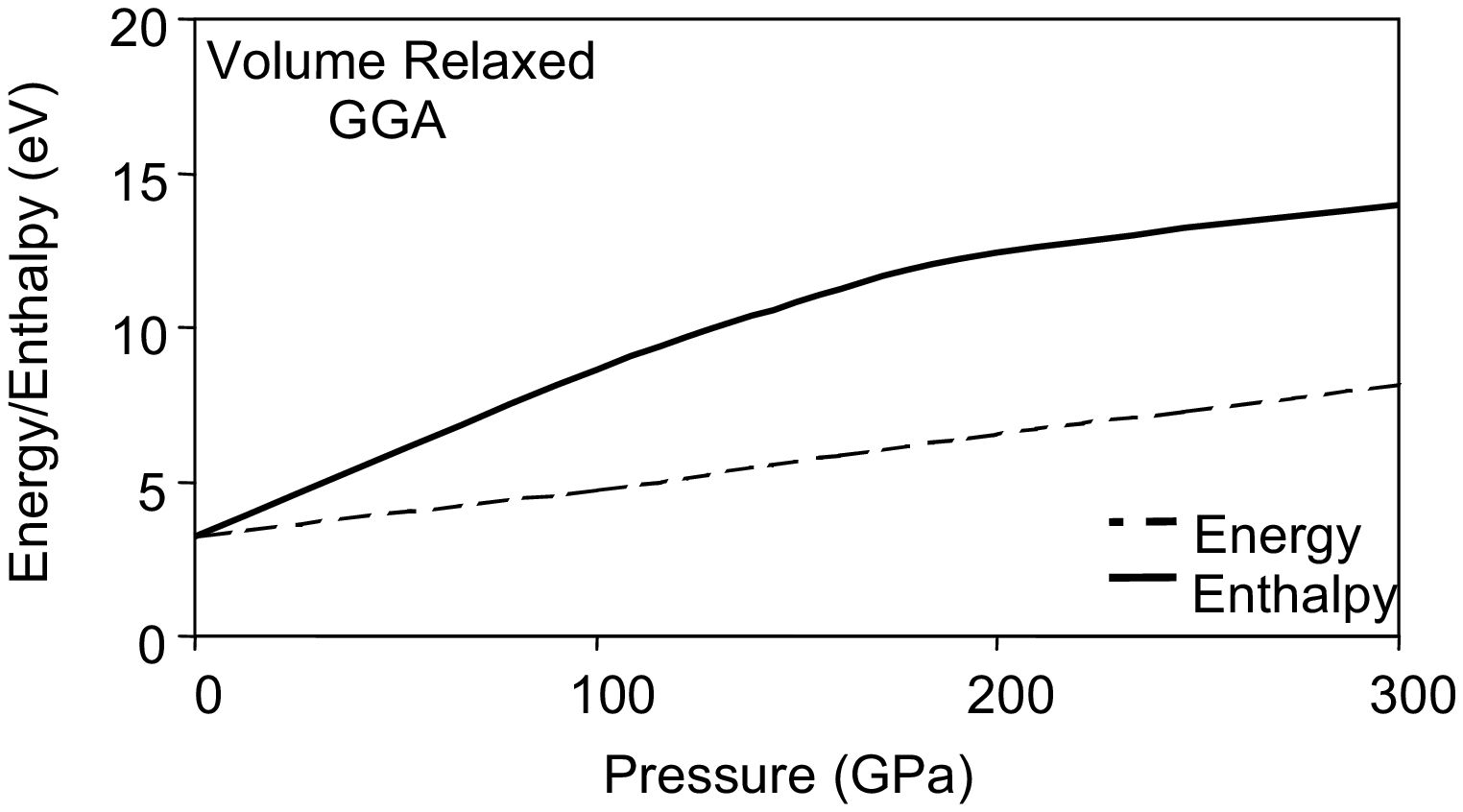,width=3.5in}}
\caption{\label{fig3}
Comparison of volume relaxed vacancy formation enthalpy and 
vacancy formation energy for a 54 atom supercell
for GGA. The vacancy formation enthalpy saturates at high pressure.} 
\end{figure}

\begin{figure}
\centerline{\epsfig{file=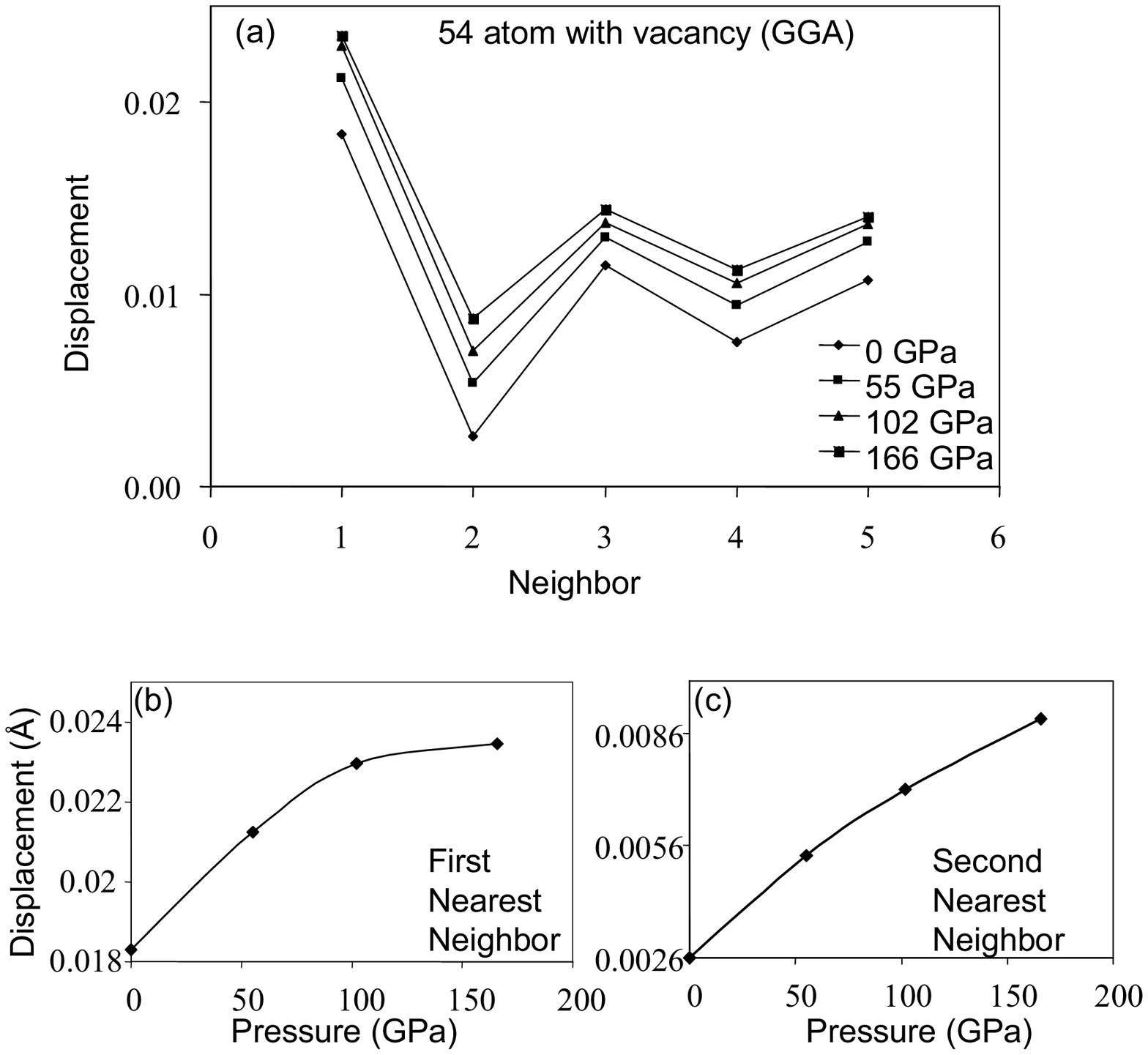,width=3.5in}}
\caption{\label{fig4}
(a)Atomic displacement around a vacancy of the neighbor atoms for the 54 
atom supercell during full (volume and structural) relaxation.
(b)First and second nearest neighbor displacements versus pressure
for the 54 atom supercell during full (volume and structural) relaxation.}
\end{figure}

\begin{figure}
\centerline{\epsfig{file=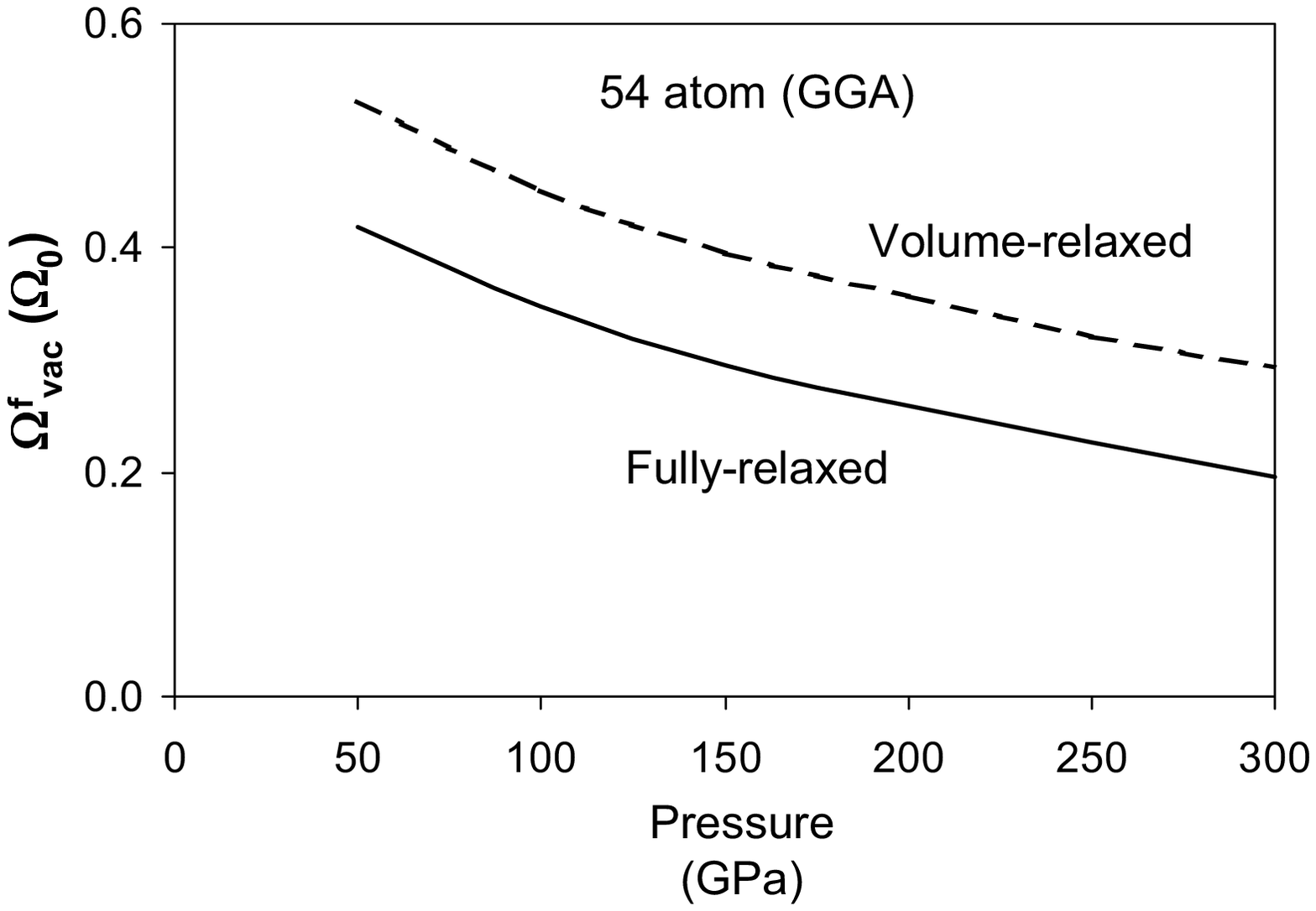,width=3.5in}}
\caption{\label{fig5}
Vacancy formation volume, $\Omega_{vac}^{f}(P)$, versus pressure. The 
vacancy formation volume is expressed relative to the volume of a single atom
in the ideal (without vacancy) Ta system, $(\Omega_0(P))$.}  
\end{figure}

\begin{figure}
\centerline{\epsfig{file=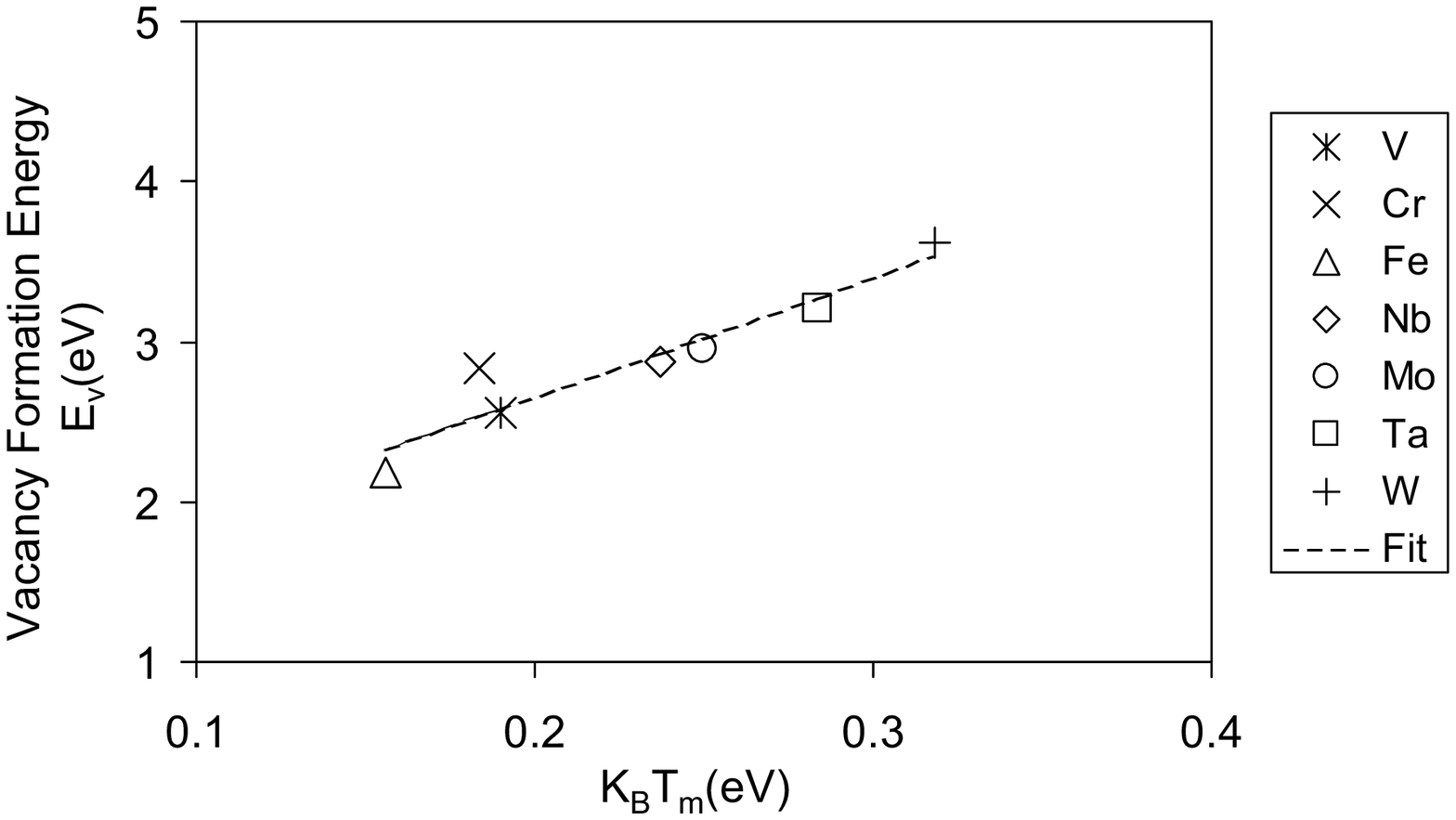,width=3.5in}}
\caption{\label{fig6}
Theoretical values \cite{moriarty} of the vacancy formation energies $E_v$
are plotted with respect to the  
melting temperatures \cite{kittel} $T_m$, for the elemental bcc transition metals.
The dashed line is obtained by a linear fit to the data. The
correlation coefficient of the fitted line is $0.886$.}  
\end{figure}

\newpage
\begin{table}
\caption{Comparison of available Ta vacancy formation energies (eV) at 
ambient pressure. The volume relaxed (vol rel) and fully relaxed (full rel)  
values for the 16 and 54 atom supercells are tabulated.
The experimental value \cite{expt} is $3.1$ eV.}
\label{table1}
\begin{tabular}{ccccccccccccccccc}
&&&16 atom&&&&&54 atom&&&&&54 atom&&&\\
&&&LDA&&&&&LDA&&&&&GGA&&&\\
&&volume&&fully&&&volume&&fully&&&volume&&fully&&\\
&&relaxed&&relaxed&&&relaxed&&relaxed&&&relaxed&&relaxed&&\\
\hline
&Ref.\onlinecite{satta} &3.29&&3.17&&&3.51&&2.99&&& &&  &&\\
&Ref. \onlinecite{moriarty}&&&&&&3.6&&3.2&&&&&3.2&&\\
&Present Work&3.26&&3.23&&&3.6&&&&&3.25&&2.95&&\\
\end{tabular}
\end{table}

\begin{table}
\caption{Total energy (GGA) in eV of the 54 atom supercell with and without vacancy 
(ideal) for different system volumes (\AA)$^3$.}
\label{table2}
\begin{tabular}{ccccccccccc}
&System Volume&&&Energy for Ideal System&&&&Energy for System with a vacancy&&\\
\hline
&953.547&&& -10237.366&&&&-10044.717&&\\
&868.987&&& -10227.952&&&&-10036.252&&\\
&789.579&&& -10204.537&&&&-10014.345&&\\
&715.163&&& -10163.550&&&&-9975.546&&\\
&645.576&&& -10100.465&&&&-9915.540&&

\end{tabular}
\end{table}

\end{document}